\documentclass[12pt]{iopart}

\expandafter\let\csname equation*\endcsname=\relax
\expandafter\let\csname endequation*\endcsname=\relax
\usepackage{graphicx}
\usepackage{mathtools}
\usepackage{hyperref}
\usepackage[normalem]{ulem}
\usepackage{amsmath}
\usepackage{amssymb}
\usepackage{amsfonts}
\usepackage{mathrsfs}
\usepackage{bm, dsfont}
\usepackage[usenames,dvipsnames]{color}
\usepackage{colortbl}
\usepackage[table]{xcolor}
\usepackage{enumitem}
\usepackage{dcolumn}
\usepackage{epstopdf}
\usepackage{cleveref}
\usepackage{hypcap}
\usepackage{verbatim, float}
\usepackage{siunitx}
\usepackage{tikz}
\usetikzlibrary{shapes.geometric, arrows}

\usepackage{xcolor}
\definecolor{navyblue}{rgb}{0.0, 0.0, 0.5}
\definecolor{orangered4}{rgb}{0.55, 0.15, 0.0}

\usepackage{listings}
\usepackage{algorithm,algorithmic}
\usepackage[algo2e,ruled,vlined]{algorithm2e}
\usepackage{caption}

\usepackage{makecell}

\usepackage{booktabs}

\usepackage{placeins}

\usepackage{adjustbox} % To adjust table size if necessary

\begin{document}
\title[]{Machine learning the Ising transition: A comparison between discriminative and generative approaches}

\author{Difei Zhang\textsuperscript{1}, Frank Sch\"afer\textsuperscript{1}, and Julian Arnold\textsuperscript{2}}

\address{$^1$ CSAIL, Massachusetts Institute of Technology, Cambridge, MA
	02139, USA}
\address{$^2$ Department of Physics, University of Basel, Klingelbergstrasse 82, CH-4056 Basel, Switzerland}
\ead{difeiz@mit.edu, franksch@mit.edu, julian.arnold@unibas.ch}
\vspace{10pt}
\begin{indented}
\item[]\today
\end{indented}

\begin{abstract}
The detection of phase transitions is a central task in many-body physics. To automate this process, the task can be phrased as a classification problem. Classification problems can be approached in two fundamentally distinct ways: through either a discriminative or a generative method. In general, it is unclear which of these two approaches is most suitable for a given problem. The choice is expected to depend on factors such as the availability of system knowledge, dataset size, desired accuracy, computational resources, and other considerations. In this work, we answer the question of how one should approach the solution of phase-classification problems by performing a numerical case study on the thermal phase transition in the classical two-dimensional square-lattice ferromagnetic Ising model.
\end{abstract}

%%%%%%%%%%%%%%%%%%%%%%%%%%%%
\section{Introduction}
\label{sec:intro}
Mapping out the phase diagram of physical systems is a central task in many-body physics~\cite{sachdev:2011,goldenfeld:2018}. Typically, this process relies heavily on human intuition and involves the identification of a few suitable low-dimensional quantities that capture the nature of each of the phases at play -- so-called order parameters -- or corresponding response functions, such as the magnetic susceptibility or the heat capacity. In recent years, physicists have begun exploring the use of machine learning to automate this task~\cite{carleo:2019,carrasquilla:2021,dawid:2022}. To this end, the task of mapping out a phase diagram can be framed as a classification task, which can be addressed in a data-driven manner using tools from machine learning~\cite{arnold:2022,arnold_prl:2024}. Based on the solution of this classification task, indicator functions can be constructed whose local maxima highlight phase boundaries (similar to traditional response functions). 

The classification tasks underlying the problem of mapping out a phase diagram can be tackled in two distinct ways~\cite{arnold_prl:2024}. In a \emph{discriminative approach}, models such as fully-connected or convolutional neural networks (NNs) are trained to approximate the probability distribution $P(y|\bm{x})$ over the labels $y \in \mathcal{Y}$ of a sample $\bm{x} \in \mathcal{X}$ given a labeled training dataset $\mathcal{D} = \{ (\bm{x}_{i},y_{i})\}_{i}$. Whereas in a \emph{generative approach}, an approximation of $P(y|\bm{x})$ is constructed via Bayes' rule 
\begin{equation}\label{eq:Bayes}
    P(y|\bm{x}) = \frac{P(\bm{x}|y)P(y)}{P(\bm{x})}
\end{equation}
based on generative models underlying the system, i.e., models for $P(\bm{x}|y)$. In general, it is unclear which of these two approaches is the most suitable choice for a given problem. In particular, this choice is expected to depend on factors such as the extent of available system knowledge, the selected model, the size of the dataset, the required accuracy, the computational budget, and the type of indicator being computed.

In this work, we tackle the question of how one most efficiently computes indicator functions to locate phase transitions in physical systems in an automated fashion. We consider two distinct settings. First, a \emph{data-driven setting} in which we lack knowledge about the underlying system in an explicit form, but configurational data of the system is available. This setting captures scenarios in which systems are only accessible experimentally. Second, a \emph{knowledge-driven setting} in which knowledge about the underlying physical system is available and can be readily exploited in the process. In particular, we assume the underlying system Hamiltonian to be known. This setting captures scenarios where systems are investigated numerically. As a physical system for our case study, we will focus on the thermal phase transition in the classical two-dimensional square-lattice ferromagnetic Ising model.

\section{Methodology}

\subsection{Detecting phase transitions using machine-learned indicators}
\label{sec:indic}
In this work, we focus on three distinct approaches for mapping out phase diagrams from data that are based on computing scalar indicators of phase transitions $I(\gamma)$ whose local maxima indicate phase boundaries~\cite{arnold_prl:2024}. For simplicity, we assume the system to be characterized by a single tuning parameter $\gamma \in \mathbb{R}$. Note that the methods have previously been extended to higher-dimensional parameter spaces~\cite{arnold_prl:2024}.  We analyze the system at a discrete set of such points $\Gamma$. The indicator computation boils down to solving classification tasks where the label $y\in \mathcal{Y}$ specifies a subset of the sampled points $\Gamma_{y}$. The three approaches differ in the specific classification tasks that are solved, i.e., in the choice of $\{ \Gamma_{y}\}_{y \in \mathcal{Y}}$. It can be shown that the indicators of all three approaches are underapproximators of the system's Fisher information~\cite{arnold_FI:2023}.

\emph{Approach 1: Supervised learning} --- The first approach, so-called supervised learning (SL)~\cite{carrasquilla:2017}, presumes some prior knowledge of the phase diagram. Here, the number of distinct phases $K$ and representative points in parameter space $\Gamma_{y}$ are assumed to be known. That is, the label $y$ plays the role of denoting each of the phases of the system where $y \in \mathcal{Y} = \{1,\dots,K \}$. The corresponding indicator can then be computed as 
\begin{align*}\label{eq:ISL}
    I_{\rm SL}(\bm{\gamma}) &= \frac{1}{K} \sum_{y \in \mathcal{Y}} \left|\frac{\partial P(y|\gamma)}{\partial \gamma}\right| ,
\end{align*}
where $P(y|\gamma) = \mathbb{E}_{\bm{x} \sim P(\cdot|\gamma)}\left[ P(y|\bm{x}) \right]$ is to be interpreted as the posterior probability of point $\gamma$ belonging to phase $y$. In this work, we have $K=2$ and always choose the leftmost and rightmost points from the sampled parameter range as our training points for phase $y=1$ and phase $y=2$, respectively.

\emph{Approach 2: Learning by confusion} --- In a second approach, so-called learning by confusion (LBC)~\cite{van:2017}, at each sampled point $\gamma\in \Gamma$, the parameter space is partitioned into two sets $\Gamma_{1}(\gamma)$ and $\Gamma_{2}(\gamma)$, where $\mathcal{Y} = \{1,2\}$, that are each comprised of the $l$ points closest to $\gamma$ in part 1 and 2 of the split parameter space, respectively. Based on these labeled sets of points, we can compute an error probability
\begin{equation}\label{eq:LBC_err}
    p_{\rm err}(\gamma) = \frac{1}{2}\sum_{y \in \{1,2 \}} \frac{1}{|\Gamma_{y}|}\sum_{\gamma' \in \Gamma_{y}}\mathbb{E}_{\bm{x} \sim P(\cdot|\gamma')} \left[p_{\rm err}(\bm{x})\right],
\end{equation}
where $p_{\rm err}(\bm{x}) = {\rm min}\left\{ P(1|\bm{x}), P(2|\bm{x}) \right\}$ is the (optimal) average error probability when predicting the label of sample $\bm{x}$. The indicator can then be computed as $I_{\rm LBC}(\gamma) =  1- 2 p_{\rm err}(\gamma)$.

\emph{Approach 3: Prediction-based method} --- In a third approach, the so-called prediction-based method (PBM)~\cite{arnold_prl:2024,schaefer:2019,arnold:2021}, each sampled value of the tuning parameter is assigned its own label $\Gamma_{y} = \{ \gamma_{y}\}, \; y \in \mathcal{Y} = \{1,\dots,|\Gamma| \}$. The indicator is then given as 
\begin{equation}\label{eq:IPBM}
    I_{\rm PBM}(\gamma) = \frac{\partial \hat{\gamma}({\gamma}) / \partial \gamma }{\sigma({\gamma})},
\end{equation}
where 
\begin{equation}
    \hat{\gamma}(\gamma) = \mathbb{E}_{\bm{x} \sim P(\cdot|\gamma)}\left[ \hat{\gamma}(\bm{x})\right] = \mathbb{E}_{\bm{x} \sim P(\cdot|\gamma)}\left[ \sum_{y \in \mathcal{Y}} P(y|\bm{x})\gamma_{y}\right]
\end{equation}
is the mean predicted value of the tuning parameter and
\begin{equation}
    \sigma(\gamma) = \sqrt{\mathbb{E}_{\bm{x} \sim P(\cdot|\gamma)}\left[ \hat{\gamma}(\bm{x})^2\right] - \left(\mathbb{E}_{\bm{x} \sim P(\cdot|\bm{\gamma})}\left[ \hat{\gamma}(\bm{x})\right]\right)^2}
\end{equation}
is the corresponding standard deviation.

\subsection{Discriminative modeling vs. generative modeling}
\label{sec:discr_gen}
In the previous section, we have shown how to reduce the problem of detecting phase transitions from data to the computation of a scalar indicator of phase transitions. Crucially, this computation involves solving classification tasks, i.e., one needs to model the probability $P(y|\bm{x})$ of a label $y \in \mathcal{Y}$ given a sample $x \in \mathcal{X}$. In the following, we will refer to approximate quantities with $\tilde{\cdot}$. Once such a model $\tilde{P}(y|\bm{x})$ is obtained, we can compute an estimate of the indicator $\tilde{I}$ by substituting $P(y|\bm{x})$ with $\tilde{P}(y|\bm{x})$ and approximating expected values with sample means. Models of this type can be obtained in two fundamentally distinct ways.

\emph{Discriminative modeling} --- In a discriminative approach, given a set of samples $\mathcal{D}_{\gamma}$ drawn from $P(\bm{x}|\gamma)$ for each $\gamma \in \Gamma_{y}$, i.e., a labeled set of data points, a model $\tilde{P}(y|\bm{x})$ approximating $P(y|\bm{x})$ is constructed directly. Throughout this work, we assume that $|\mathcal{D}_{\gamma}|$ is the same $\forall \gamma \in \Gamma$. Typically, $\tilde{P}(y|\bm{x})$ is parametric, i.e., $\tilde{P}_{\bm{\theta}}(y|\bm{x})$ is represented as an NN whose parameters $\bm{\theta}$ are optimized in a supervised fashion to solve the classification task. That is, one minimizes a cross-entropy loss function
\begin{equation}\label{eq:loss1}
    \mathcal{L}(\bm{\theta}) = - \frac{1}{|\mathcal{Y}|} \sum_{y \in \mathcal{Y}} \frac{1}{|\mathcal{D}_{y}|} \sum_{\bm{x} \in \mathcal{D}_{y}}  {\rm ln}\left(\tilde{P}_{\bm{\theta}}(y|\bm{x})\right),
\end{equation}
where $\mathcal{D}_{y} = \{ \bm{x} \in \mathcal{D}_{\gamma} |\gamma \in \Gamma_{y}\}$. For approach 3, it can be useful to use a parametric predictive model $\hat{\gamma}_{\bm{\theta}}(\bm{x})$ that is trained to solve a regression task instead of a classification task. Here, we use a mean-square-error loss
\begin{equation}\label{eq:loss2}
    \mathcal{L}(\bm{\theta}) = \frac{1}{|\mathcal{Y}|} \sum_{y \in \mathcal{Y}} \frac{1}{|\mathcal{D}_{y}|} \sum_{\bm{x} \in \mathcal{D}_{y}}  \left(\hat{\gamma}_{\bm{\theta}}(\bm{x}) - \gamma_{y} \right)^2.
\end{equation}
We will use this strategy throughout this work for approach 3. As discriminative models, we use simple feedforward NNs, i.e., multilayer perceptrons (MLPs) and convolutional NNs (CNNs) where fitting. Details on the choice of architecture, hyperparameters, and training are summarized in~\ref{abbr:training_details}.

\emph{Generative modeling} --- To solve the underlying classification tasks in a generative manner, we can choose $P({\gamma}|y) = 1 / |\Gamma_{y}|$ for $\gamma\in \Gamma_{y}$ and zero otherwise. This yields $P(\bm{x}|y) = \sum_{\gamma \in \Gamma} P(\bm{x}|\gamma) P(\gamma|y) = \frac{1}{|\Gamma_{y}|}\sum_{\gamma \in \Gamma_{y}} P(\bm{x}|\gamma)$. Using Bayes' rule [Eq.~\eqref{eq:Bayes}] with $P(y)=1/|\mathcal{Y}|$, we have
\begin{equation}\label{eq:Bayes2}
    P(y|\bm{x}) =  \frac{\frac{1}{|\Gamma_{y}|}\sum_{\gamma \in \Gamma_{y}} P(\bm{x}|\gamma)}{\sum_{y' \in \mathcal{Y}}\frac{1}{|\Gamma_{y'}|}\sum_{\gamma' \in \Gamma_{y'}} P(\bm{x}|\gamma')}.
\end{equation}
Based on this expression, we can obtain models $\tilde{P}(y|\bm{x})$ given models for $P(\bm{x}|\gamma)$ -- the probability distributions underlying the measurement statistics at various discrete points in parameter space. Note that we want to be able to efficiently sample from these distributions. 

Non-parametric models construct an approximation $\tilde{P}(\bm{x}|\gamma)$ from data without using free tunable parameters. In this work, we will consider such models obtained via histogram binning $\tilde{P}(\bm{x}|\gamma) = 1/|\mathcal{D}_{\gamma}|\sum_{\bm{x'} \in \mathcal{D}_{\gamma}} \delta_{\bm{x},\bm{x}'}$. While simple (and asymptotically unbiased), such models are expected to perform poorly if the state space $\mathcal{X}$ gets too large. This is particularly problematic in our case because the state space of many-body systems typically grows exponentially with the system size and large system sizes are needed to observe collective phenomena such as phase transitions.

More generally, parametric models $\tilde{P}_{\bm{\theta}}(\bm{x}|\gamma)$ can be constructed. To illustrate the training procedure, in the following, we will omit the conditioning on $\gamma$. Ideally, we would like $P = \tilde{P}_{\bm{\theta}}$. The Kullback-Leibler (KL) divergence,
\begin{equation}
    {\rm KL}(P|\tilde{P}_{\theta}) = \sum_{\bm{x} \in \mathcal{X}} \tilde{P}_{\bm{\theta}}(\bm{x}) \ln\left(\frac{\tilde{P}_{\bm{\theta}}(\bm{x})}{P(\bm{x})}\right),
\end{equation}
allows us to measure how different the two distributions are. In particular, ${\rm KL}(P|\tilde{P}_{\bm{\theta}})=0$ if the two distributions match and ${\rm KL}(P|\tilde{P}_{\bm{\theta}})>0$ otherwise. Note that the KL divergence is not symmetric, and we have
\begin{equation}
    {\rm KL}(\tilde{P}_{\bm{\theta}}|P) = \sum_{\bm{x} \in \mathcal{X}} P(\bm{x}) \ln\left(\frac{P(\bm{x})}{\tilde{P}_{\bm{\theta}}(\bm{x})}\right).
\end{equation}
However, also in this case ${\rm KL}(\tilde{P}_{\bm{\theta}}|P)=0$ holds if the two distributions match. So \emph{a priori}, both represent valid objectives to optimize.

Using ${\rm KL}(\tilde{P}_{\bm{\theta}}|P)$, the so-called \textit{forward} KL divergence, we have
\begin{equation}
    {\rm KL}(\tilde{P}_{\bm{\theta}}|P) \approx \frac{1}{|\mathcal{D}|}\sum_{\bm{x} \in \mathcal{D}} \ln\left(\frac{P(\bm{x})}{\tilde{P}_{\bm{\theta}}(\bm{x})}\right),
\end{equation}
where $\mathcal{D}$ is a dataset of points sampled from $P$. Based on this, we can formulate a loss function
\begin{equation}
    \mathcal{L}(\bm{\theta}) = - \frac{1}{|\mathcal{D}|}\sum_{\bm{x} \in \mathcal{D}}\ln \left(\tilde{P}_{\bm{\theta}}(\bm{x})\right).
\end{equation}
Note that minimizing this loss is equivalent to maximizing the log-likelihood, i.e., performing maximum-likelihood estimation.
\begin{figure}[bth!]
\includegraphics[width=0.8\linewidth]{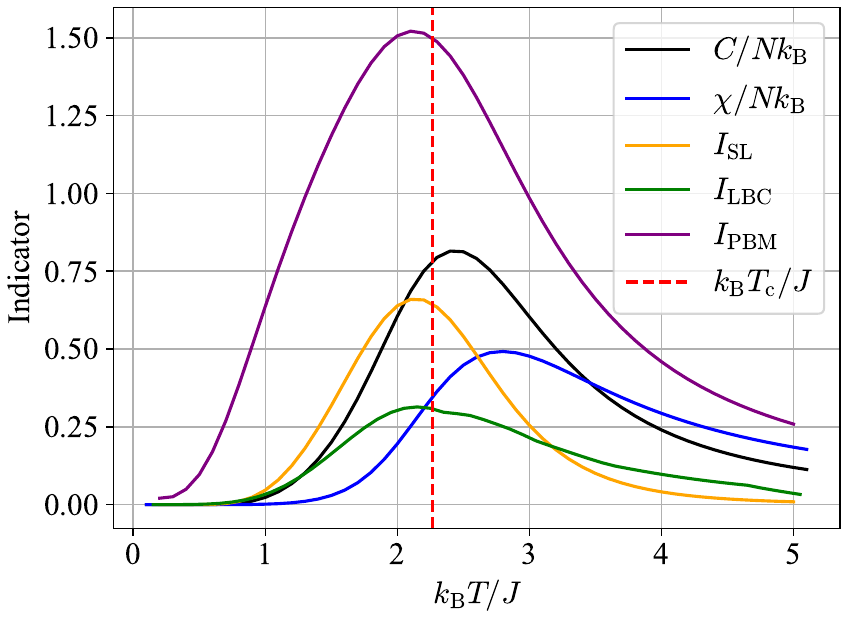}
\centering
\caption{Physical and ground-truth indicators (corresponding to Bayes-optimal predictive models) of the three ML methods for the Ising model [Eq.~\eqref{eq:IsingH}, $L=4$]. For the physical indicators we consider the heat capacity $C/Nk_{\rm B}$ and the magnetic susceptibility $\chi/Nk_{\rm B}$, where $N=L^2=4^2=16$ is the number of spins. For the machine-learning methods, we consider supervised learning (SL), learning by confusion (LBC; $l=5$), and the prediction-based method (PBM). Throughout this article, the set $\Gamma$ is composed of a uniform grid with 51 points ranging from $0.1$ $k_{\rm B}T/J$ to $5.1$ $k_{\rm B}T/J$. The ground-truth indicator is computed from $\{P(\cdot |\gamma) \}_{\gamma \in \Gamma}$ which can be computed exactly for this lattice size. The critical point given by Onsager's solution~\cite{onsager:1944} is highlighted by the red dashed line.}
\label{fig:example_indicators}
\end{figure}
If we use the \emph{reverse} KL divergence instead, we have
\begin{equation}
    {\rm KL}(P|\tilde{P}_{\bm{\theta}}) =  \sum_{\bm{x}} \tilde{P}_{\bm{\theta}}(\bm{x}) \ln \left(\frac{\tilde{P}_{\bm{\theta}}(\bm{x})}{P(\bm{x})}\right).
\end{equation}
In case the system is described by a Boltzmann distribution ${\rm P}(\bm{x}) = e^{-\beta H(\bm{x})}/Z$ with $Z$ being the partition function, $H$ the system Hamiltonian, and $\beta = 1/k_{\rm{B}}T$ the inverse temperature, we can write   
\begin{equation}
    {\rm KL}(P|\tilde{P}_{\bm{\theta}}) =\sum_{\bm{x} \in \mathcal{X}} \tilde{P}_{\bm{\theta}}(\bm{x})  \left[\ln\left(\tilde{P}_{\bm{\theta}}(\bm{x})\right) + \beta H(\bm{x}) + \ln(Z)\right].
\end{equation}
Thus, the relevant loss function is
\begin{equation}
    \mathcal{L}(\bm{\theta}) =\sum_{\bm{x} \in \mathcal{X}} \tilde{P}_{\bm{\theta}}(\bm{x})  \left[\ln \left( \tilde{P}_{\bm{\theta}}(\bm{x}) \right) + \beta H(\bm{x})\right],
\end{equation}
which can be approximated solely by drawing samples from $\tilde{P}_{\bm{\theta}}$. In general, we can use it whenever the target density $P$ is known up to a normalization constant, i.e., given that the Hamiltonian is known because any probability distribution can be modeled as a Boltzmann distribution with a suitable Hamiltonian~\cite{huembeli:2022,dawid:2024}.

A common approach to designing generative models with explicit, tractable densities in discrete state spaces is to leverage the chain rule of probability. This involves expressing the joint probability of all state variables, $\bm{x} = (x_{1}, \dots, x_{N})$, as a product of conditional probabilities
\begin{equation}
    \tilde{P}_{\bm{\theta}}(\bm{x}) = \Pi_{i=1}^{N} \tilde{P}_{\bm{\theta}}(x_{i}|x_{1},\dots,x_{i-1}).
\end{equation}
If the factors are parametrized by NNs, such an architecture is referred to as an autoregressive NN. Given that the Ising model we are going to study lives on a two-dimensional lattice, we opt for the so-called \emph{PixelCNN}~\cite{van:2016} which uses convolutional filters and has previously been shown to accurately describe the Ising model~\cite{wu:2019,nicoli:2020,wu:2021}. Figure~\ref{fig:pixelcnn_reproduce} shows that the PixelCNN can accurately reproduce physical quantities, such as the energy or magnetization of the Ising model. Details on the choice of architecture, hyperparameters, and training are summarized in~\ref{abbr:training_details}.

\subsection{Ising model}
\label{sec:Ising}

As an example for our case study, we consider the two-dimensional square-lattice ferromagnetic Ising model described by the following Hamiltonian
\begin{equation}\label{eq:IsingH}
 H(\bm{\sigma}, \, J) = -  J\sum_{\langle ij\rangle} \sigma_{i}\sigma_{j}.
\end{equation}
Here, the sum runs over all nearest-neighboring sites (with periodic boundary conditions), $J$ is the interaction strength $(J>0)$, and $\sigma_{i} \in \{+1,-1 \}$ denotes the discrete spin variable at lattice site $i$. This results in a state space of size $2^{L \times L}$ for a square lattice of linear size $L$. The system is completely characterized by its spin configuration $\bm{\sigma} = (\sigma_{1},\sigma_{2},\dots,\sigma_{L \times L})$. We consider the system to be in thermal equilibrium, i.e., the distribution of spin configurations is assumed to be Boltzmann 
\begin{equation}
    P(\bm{\sigma}|\gamma) = e^{-\beta H(\bm{\sigma},\; J)}/Z
\end{equation}
with $Z$ being the partition function, inverse temperature $\beta = 1/k_{\rm{B}}T$, and $\gamma = k_{B}T/J$. We draw spin configurations from Boltzmann distributions at distinct temperatures via Markov chain Monte Carlo (MCMC), see~\ref{app:data_generation} for details.

This system exhibits a phase transition between a paramagnetic (disordered) phase at high temperature $T> T_{\rm c}$ and a ferromagnetic (ordered) phase at low temperature $T < T_{\rm c}$, where the critical temperature $T_{\rm c}$ obtained in the thermodynamic limit is ${2J}/{k_{\rm B} \ln(1+\sqrt{2})}$. Figure~\ref{fig:example_indicators} shows how well the machine-learning methods described above highlight the phase transition in comparison to physical quantities, such as the heat capacity or magnetic susceptibility.

For such Boltzmann distributions, it is known that the energy is a minimum sufficient statistic for the parameter $\gamma$~\cite{arnold:2022,arnold_prl:2024}. Thus, to detect phase transitions using machine-learned indicators, it suffices to consider the energy, i.e., $H$ [Eq.~\eqref{eq:IsingH}], instead of the spin configuration $\bm{\sigma}$ itself. This corresponds to an optimal lossless compression of the state space and enables the simple non-parametric modeling approach described in the previous section. Note, however, that this requires knowledge of the underlying Hamiltonian, i.e., it can only be performed in the knowledge-driven setting.

\begin{figure}[htb!]
\includegraphics[width=\linewidth]{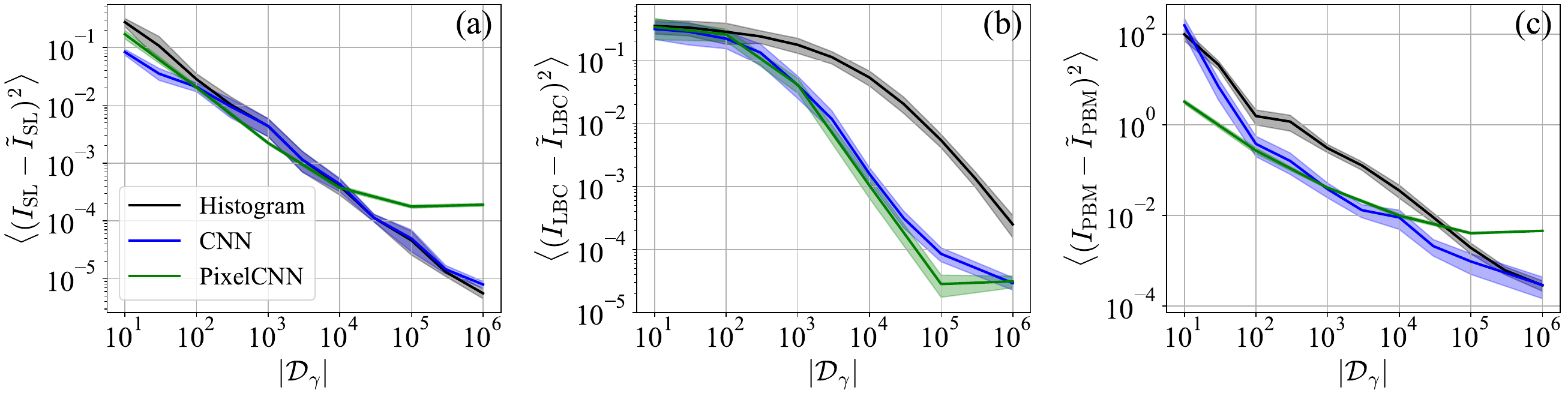}
%original figure: results/error-datasetsize-senario1.png
\centering
\caption{Mean-square-error between machine-learned indicator $\tilde{I}$ and ground-truth indicator $I$ as a function of the dataset size $|\mathcal{D}_{\gamma}|$ for the Ising model [Eq.~\eqref{eq:IsingH}, $L=4$] in the data-driven setting. We consider the indicators of (a) supervised learning (SL), (b) learning by confusion (LBC; $l=5$), and (c) prediction-based method (PBM). Shaded band corresponds to the 1-standard deviation error estimated from 5 independent runs.}
\label{fig:datadriven_error_datasetsize}
\end{figure}

\section{Results: Data-driven setting}
\label{sec:data}
First, we consider a data-driven setting in which only raw snapshots of the spin configuration are available but no information about the underlying Hamiltonian. In this scenario, the key quantity measuring the associated cost is the number of data points being collected.

\begin{figure}[htb!]
\includegraphics[width=\linewidth]{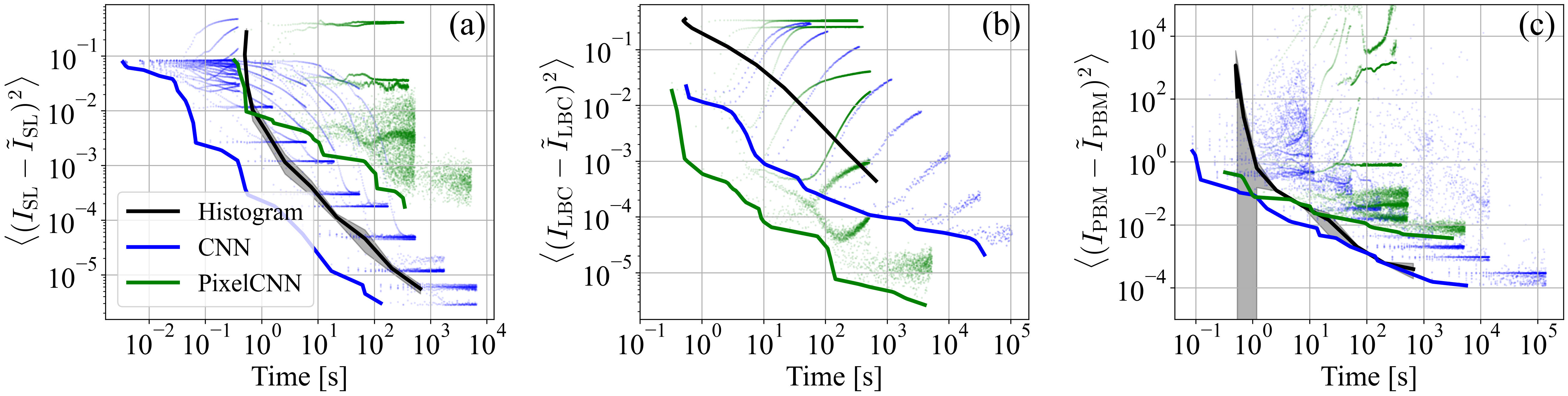}
% original figure: results/error-computationtime-senario1(noMCMC).png
\centering
\caption{Mean-square-error between machine-learned indicator $\tilde{I}$ and ground-truth indicator $I$ as a function of the computation time for the Ising model [Eq.~\eqref{eq:IsingH}, $L=4$] in the data-driven setting. Here, the computation time encompasses both the time to construct/train the relevant models as well as the subsequent indicator computation. The time required for generating the data is not counted. We consider the indicators of (a) supervised learning (SL), (b) learning by confusion (LBC; $l=5$), and (c) prediction-based method (PBM). Shaded bands correspond to the 1-standard deviation error estimated from 5 independent runs. Scattered points show optimization runs with distinct hyperparameter settings, see~\ref{abbr:training_details}. Bold lines highlight their envelope, i.e., the hyperparameter setting achieving the lowest error at a given computation time budget.}
\label{fig:datadriven_error_comptime_noMCMC}
\end{figure}

Figure~\ref{fig:datadriven_error_datasetsize} shows the error between machine-learned indicator $\tilde{I}$ and ground-truth indicator $I$ (see Fig.~\ref{fig:example_indicators}) as a function of the dataset size $|\mathcal{D}_{\gamma}|$. We observe that both the NN-based discriminative and generative approaches can outperform the naive histogram-based generative approach, i.e., achieve a lower error for a given number of samples. In all cases, this advantage seems to diminish as the number of samples becomes large in comparison to the size of the state space $|\mathcal{X}|=2^{4\times 4} = 65536$. In this regime, the histogram-based approach becomes highly accurate whereas the accuracy of NN-based methods starts to suffer from internal biases. In the case of supervised learning, almost no advantage can be observed, this is because in this case, the optimal prediction is solely based on estimating the probability of drawing a ground-state configuration~\cite{arnold:2022} which can be estimated with very few samples using naive histogram binning. LBC and PBM are instead sensitive to the entire state space. It is also interesting to observe that NN-based generative modeling can be \emph{on par or even more efficient} compared to the discriminative approach. This also holds when considering models with a different number of trainable parameters, see Fig.~\ref{fig:smaller_model_LBC}(a) in the Appendix.

The advantage of the NN-based methods is even more striking in Fig.~\ref{fig:datadriven_error_comptime_noMCMC} which shows the error as a function of the computation time. For SL and PBM, at a fixed computation time budget, the NN-based discriminative approach yields the most accurate approximations of the ground-truth indicators. For LBC, the NN-based generative approach is most favorable. Note that the only factor influencing the computation time in the nonparametric generative approach is the chosen dataset size. In contrast, the performance of the parametric approaches can differ strongly depending on the choice of hyperparameters, such as the number of training epochs. The cost of the hyperparameter tuning is not explicitly included here. Instead, we show different optimization runs with distinct hyperparameter settings (scattered points in Fig.~\ref{fig:datadriven_error_comptime_noMCMC}). The envelopes marking the hyperparameter settings that achieve the lowest error at a given computation time budget are highlighted by bold lines.

\section{Results: Knowledge-driven setting}
\label{sec:knowledge}

In the knowledge-driven setting, the underlying Hamiltonian is assumed to be known. However, raw data is unavailable and must be generated through simulation. Therefore, we include the data generation in the overall computation time. In this setting, the key quantity measuring the cost is the computation time.

\begin{figure}[bth!]
%original: results/error-computationtime-2.png
\includegraphics[width=\linewidth]{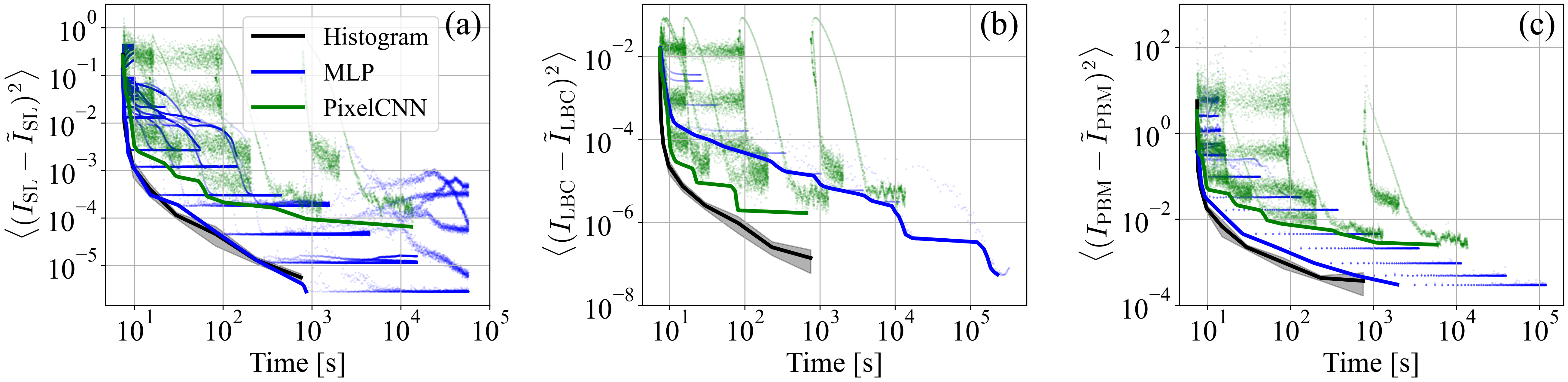}
\centering
\caption{Mean-square-error between machine-learned indicator $\tilde{I}$ and ground-truth indicator $I$ as a function of the computation time for the Ising model [Eq.~\eqref{eq:IsingH}, $L=4$] in the knowledge-driven setting. Here, the time spent for generating the data via MCMC is included in the overall computation time. We consider the indicators of (a) supervised learning (SL), (b) learning by confusion (LBC; $l=5$), and (c) prediction-based method (PBM)). Shaded bands correspond to the 1-standard deviation error estimated from 5 independent runs. Scattered points show optimization runs with distinct hyperparameter settings, see~\ref{abbr:training_details}. Bold lines highlight their envelope, i.e., the hyperparameter setting achieving the lowest error at a given computation time budget.}
\label{fig:knowledgedriven_error_comptime_complete}
\end{figure}

Figure~\ref{fig:knowledgedriven_error_comptime_complete} shows the error between machine-learned indicator $\tilde{I}$ and ground-truth indicator $I$ as a function of the computation time. For all three approaches, the nonparametric generative method using the sufficient statistic yields the best performance at a given computation time except for very small computation time budgets, i.e., very high error tolerances. This is because the number of unique energies, $|\mathcal{X}_{E}| = L^2-1$ (for even $L$), is much smaller than the number of unique configurations, $|\mathcal{X}| = 2^{L^2}$. Hence, histogram binning results in accurate results for all but the smallest number of samples. In the knowledge-driven setting, all data-driven strategies are also available. However, as expected, the data-driven strategies perform worse than the knowledge-driven strategies, see Fig.~\ref{fig:knowledgedriven_error_comptime_comparison} in~\ref{app:additional_results}. 

Examining how the error decreases as a function of the dataset size in the knowledge-driven setting, as shown in Fig. \ref{fig:knowledgedriven_error_datasetsize} in~\ref{app:additional_results}, both generative and discriminative approaches perform similarly. The NN-based generative approach is slightly favored for smaller number of samples and the other two approaches are favored for larger number of samples, $|\mathcal{D}_{\gamma}| > 10^4$. Again, this is also observed to hold when considering models with a different number of trainable parameters, see Fig.~\ref{fig:smaller_model_LBC}(b) in the Appendix.

\section{Conclusion and outlook}
\label{sec:concl}
In this work, we have investigated different approaches for solving the classification task underlying popular machine-learning methods for detecting phase transitions in an automated fashion on the example of the two-dimensional square-lattice ferromagnetic Ising model. 

In the data-driven setting, we find that NN-based approaches are favored over the nonparametric approach in terms of computation time. When only having access to a small number of samples $|\mathcal{D}_{\gamma}|\lessapprox|\mathcal{X}| = 2^{L^2}$, NN-based methods also achieve lower errors compared to the nonparametric approach. We find that the NN-based generative approach is most favored in combination with LBC, whereas the NN-based discriminative approach is most favored in combination with SL and PBM. The efficiency of discriminative LBC may be further improved using multi-tasking, see Ref.~\cite{arnold_fast:2023}. In this case, the NN-based discriminative approach becomes comparable with the NN-based generative approach. However, the efficiency of the training for parametric generative models may also be improved along similar lines. In Ref.~\cite{rende:2024}, for example, it was shown that accurate generative models across a wide range of the parameter space can be obtained by fine-tuning a model that is pre-trained at the critical point. In Ref.~\cite{moss:2024}, more accurate generative models were obtained by a two-stage training process consisting of a data-driven training stage based on forward KL followed by a knowledge-driven training based on Ritz's variational principle. This constitutes a promising line of research for future works.

In the knowledge-driven setting, the nonparametric approach is favored in terms of computation time for all but the smallest number of samples $|\mathcal{D}_{\gamma}| \gtrapprox |\mathcal{X}_{E}| = L^2-1$ or in case of large error tolerances, where $\mathcal{X}_{E}$ is the set of all possible energies for spin configurations in $\mathcal{X}$.

While we only considered a small system size $L=4$ in this work due to limited computational resources, we expect the aforementioned trends to hold even for larger system sizes with the size of the corresponding state space being the relevant threshold value in terms of dataset size. Similarly, we expect our conclusions to apply to a variety of different systems beyond the ferromagnetic Ising model on a two-dimensional square lattice, particularly thermal transitions in classical lattice models with discrete degrees of freedom at thermal equilibrium for which the same methods can be readily applied. In future work, it will be of interest to investigate to what extent these findings apply to the quantum domain and nonequilibrium settings, where different generative models may be required, and exact sufficient statistics are typically unavailable.

\section*{Acknowledgment}
We would like to thank Chris Rackauckas, Alan Edelman, Niels L\"orch, Flemming Holtorf, Petr Zapletal, Kim Nicoli, Christoph Bruder, and George Barbastathis for helpful discussions.
The authors acknowledge the MIT SuperCloud and Lincoln Laboratory Supercomputing Center for providing HPC resources that have contributed to the research results reported within this paper. 
We acknowledge financial support from the MIT-Switzerland Lockheed Martin Seed Fund and MIT International Science and Technology Initiatives (MISTI). 
J.A. acknowledges financial support from the Swiss National Science Foundation individual grant (grant no. 200020 200481).
This material is based upon work supported by the U.S. National Science Foundation under award Nos CNS-2346520, PHY-2028125, RISE-2425761, DMS-2325184, OAC-2103804, and OSI-2029670, by the Defense Advanced Research Projects Agency (DARPA) under Agreement No. HR00112490488,  by the Department of Energy, National Nuclear Security Administration under Award Number DE-NA0003965 and by the United States Air Force Research Laboratory under Cooperative Agreement Number FA8750-19-2-1000.  Neither the United States Government nor any agency thereof, nor any of their employees, makes any warranty, express or implied, or assumes any legal liability or responsibility for the accuracy, completeness, or usefulness of any information, apparatus, product, or process disclosed, or represents that its use would not infringe privately owned rights. Reference herein to any specific commercial product, process, or service by trade name, trademark, manufacturer, or otherwise does not necessarily constitute or imply its endorsement, recommendation, or favoring by the United States Government or any agency thereof. The views and opinions of authors expressed herein do not necessarily state or reflect those of the United States Government or any agency thereof. The views and conclusions contained in this document are those of the authors and should not be interpreted as representing the official policies, either expressed or implied, of the United States Air Force or the U.S. Government.

\section*{Data availability statement}
The code for reproducing the findings of this study is openly available~\cite{code}.

\appendix

\section{Details on data generation}\label{app:data_generation}
The Ising dataset was generated using the Metropolis-Hastings algorithm for each temperature parameter. In each update step, all $4\times4$ spins are traversed randomly. We set the thermalization period to $10^5$ steps, after which we assume the Markov Chain has reached its stationary distribution, which matches the Boltzmann distribution at the given temperature. To minimize correlations between adjacent states, we sampled the Markov Chain every $10$ steps, ultimately collecting $10^6$ spin configurations to create our largest dataset. Further, for the knowledge-driven settings, we also calculated the energy of each spin configuration, which led to another $10^6$ dataset consisting of scalar energies. Those two datasets serve as the training foundation for all models.

\begin{table}[htb!]
\centering
\begin{adjustbox}{max width=\textwidth}
\begin{tabular}{|l|c|c|c|c|c|c|}
\hline
 & \textbf{SL-Data} & \textbf{LBC-Data} & \textbf{PBM-Data} & \textbf{SL-Knowledge} & \textbf{LBC-Knowledge} & \textbf{PBM-Knowledge} \\
\hline
\multicolumn{7}{|c|}{\textbf{Architecture Details}} \\
\hline
Model Type & CNN  & CNN  & CNN & MLP & MLP & MLP \\
Number of Hidden Layers & 3 & 3 & 3 & 3 & 3 & 3 \\
Neurons per Layer (MLP) & 10 & 10 & 10 & 32 & 32 & 32 \\
Filter Size (CNN) & 3 $\times$ 3 & 3 $\times$ 3 & 3 $\times$ 3 & N/A & N/A & N/A \\
% Pooling Type (CNN) & Max & Avg & Max & N/A & N/A & N/A \\
Activation Function & ReLU & ReLU & ReLU & ReLU & LeakyReLU & ReLU \\
Total Number of Parameters & 3894 & 3894 & 3883 & 2242 & 2242 & 2209 \\
\hline
\multicolumn{7}{|c|}{\textbf{Training Hyperparameters}} \\
\hline
Learning Rate & 0.001 & 0.001 & 0.001 & 0.001 & 0.001 & 0.001 \\
Batch Size & 64 & 64 & 64 & 64 & 64 & 64 \\
Optimizer & Adam & Adam & Adam & Adam & Adam & Adam \\
% Dropout Rate & 0.5 & 0.3 & 0.4 & 0.2 & 0.5 & 0.3 \\
% Regularization & L2 (0.01) & L1 (0.01) & None & L2 (0.001) & L2 (0.01) & None \\
% Weight Initialization & He Normal & He Normal & He Normal & He Normal & He Normal & He Normal \\
Number of Epochs & 100 & 100 & 100 & 1000 & 1000 & 1000 \\
\hline
\end{tabular}
\end{adjustbox}
\caption{Descriptions of the six NN-based discriminative models utilized in this work. We distinguish between the three different methods for detecting phase transitions (SL, LBC, and PBM), as well as the setting (data-driven vs. knowledge-driven).}
\label{tab:nn_descriptions_discriminative}
\end{table}

\section{Details on training procedure}\label{abbr:training_details}
For the non-parametric generative approach using histogram binning, we use the spin configuration dataset for data-driven settings and the energy dataset for knowledge-driven settings. In each setting, we construct the corresponding empirical distribution, which serves as the underlying data distribution. These empirical distributions then serve as input to the three approaches for detecting phase transitions from data. We perform five independent runs to assess the variability of each estimation around the average.

\begin{table}[h!]
\centering
\begin{adjustbox}{max width=\textwidth}
\begin{tabular}{|l|c|c|}
\hline
 & \textbf{Data-Driven} & \textbf{Knowledge-Driven} \\
\hline
\multicolumn{3}{|c|}{\textbf{Architecture Details}} \\
\hline
Model Type & PixelCNN  & PixelCNN  \\
Number of Hidden Layers & 3 & 3 \\
Neurons per Layer (MLP) & 32 & 32\\
Filter Size (CNN) & 3 $\times$ 3 & 3 $\times$ 3 \\
% Pooling Type (CNN) & Max & Avg & Max & N/A & N/A & N/A \\
Activation Function & PReLU & PReLU \\
Total Number of Parameters & 2017 & 2017 \\
\hline
\multicolumn{3}{|c|}{\textbf{Training Hyperparameters}} \\
\hline
Batch Size & 64 & 64 \\
Learning Rate & 0.001, 0.0001 & 0.01  \\
Optimizer & Adam & Adam \\
Scheduler & N/A & ReduceLROnPlateau(factor=0.92, patience=100, min\_lr=1e-6)\\
Regularization & N/A & Annealing learning rate\\
% Weight Initialization & He Normal & He Normal \\
Number of Epochs \& Annealing Rate & (10, N/A), (100, N/A), (1000, N/A), (10000, N/A) & (100, 0.9), (1000, 0.98), (1000, 0.99), (10000, 0.998) \\
\hline
\end{tabular}
\end{adjustbox}
\caption{Descriptions of the two NN-based generative models utilized in this work. All computations are run on a V-100 GPU.}
\label{tab:nn_descriptions_generative}
\end{table}

For the discriminative approach, we employed multilayer perceptrons (MLPs) in the knowledge-driven setting and convolutional neural networks (CNNs) for the data-driven setting. The architectures and hyperparameters for both MLPs and CNNs are detailed in Tab.~\ref{tab:nn_descriptions_discriminative}. The models were carefully designed to ensure comparable architectures, including the same number of parameters, layers, activation functions, and other key elements, allowing for a fair performance comparison. After conducting five independent runs on a V-100 GPU, we calculated the average performance and the variability around the mean.

For the parametric generative approach, we chose the PixelCNN architecture, which is an autoregressive model with an explicit, tractable density. In the data-driven setting, we utilized the spin configurations dataset with the forward KL divergence as the loss function. For the knowledge-driven setting, instead of using the energy dataset, we assumed knowledge of how to compute the Hamiltonian and employed the reverse KL divergence as the loss function. Notably, careful design of the annealing rate was essential in this approach. A detailed list of hyperparameter settings can be found in Tab.~\ref{tab:nn_descriptions_generative}.

\section{Additional results}\label{app:additional_results}
\begin{figure}[htb!]
\includegraphics[width=\linewidth]{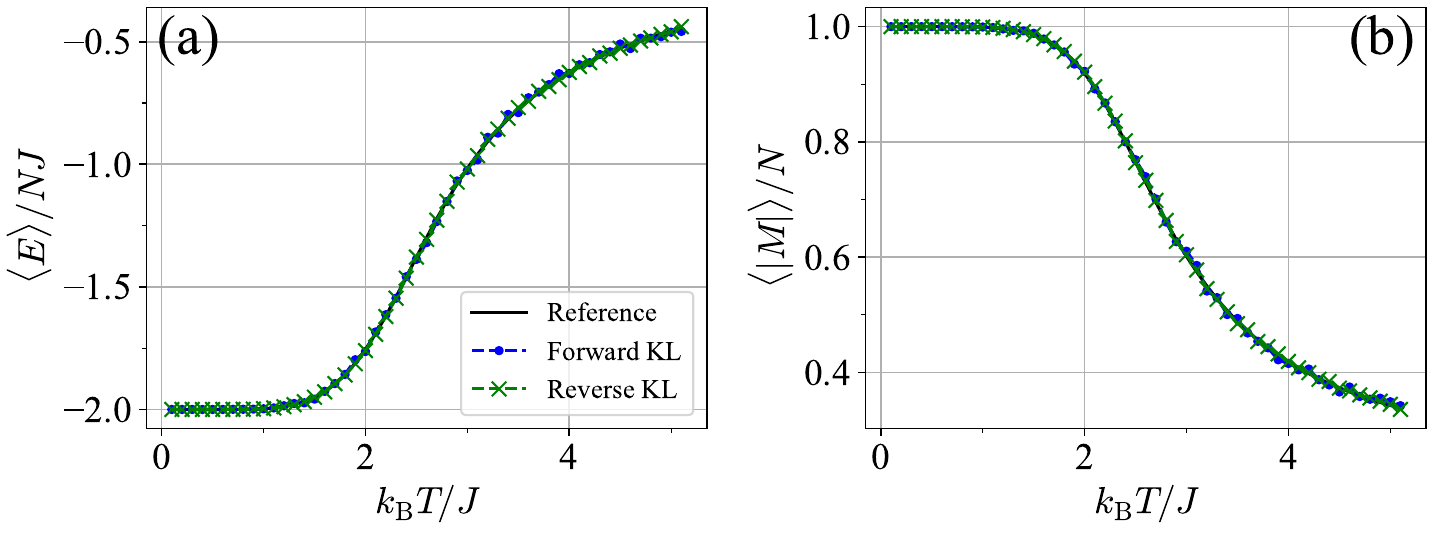}
\centering
\caption{(a) Mean energy and (b) mean magnetization of the $4 \times 4$ Ising model estimated from trained PixelCNNs using either the forward KL divergence (data-driven) or reverse KL divergence (knowledge-driven) as an objective. Here, in the data-driven setting, we utilize $|\mathcal{D}_{\gamma}|=10^6$ and train for 10 epochs. In the knowledge-driven setting, we train for 10'000 epochs. We obtain the most accurate results for these settings, i.e., these settings correspond to the most resources being utilized. For all other hyperparameters, see Tab.~\ref{tab:nn_descriptions_generative}.}
\label{fig:pixelcnn_reproduce}
\end{figure}

\begin{figure}[htb!]
\includegraphics[width=\linewidth]{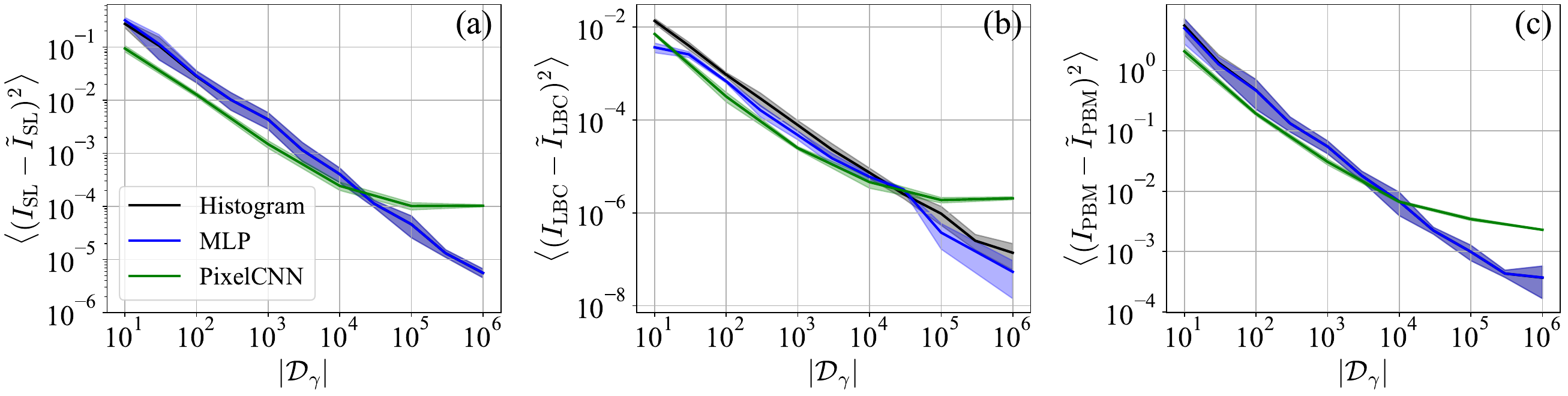}
\centering
\caption{Mean-square-error between machine-learned indicator $\tilde{I}$ and ground-truth indicator $I$ as a function of the dataset size $|\mathcal{D}_{\gamma}|$ for the Ising model [Eq.~\eqref{eq:IsingH}, $L=4$] in the knowledge-driven setting. We consider the indicators of (a) supervised learning (SL), (b) learning by confusion (LBC; $l=5$), and (c) prediction-based method (PBM). Shaded band corresponds to the 1 standard deviation error estimated from 5 independent runs.}
\label{fig:knowledgedriven_error_datasetsize}
\end{figure}

\begin{figure}[htb!]
\includegraphics[width=\linewidth]{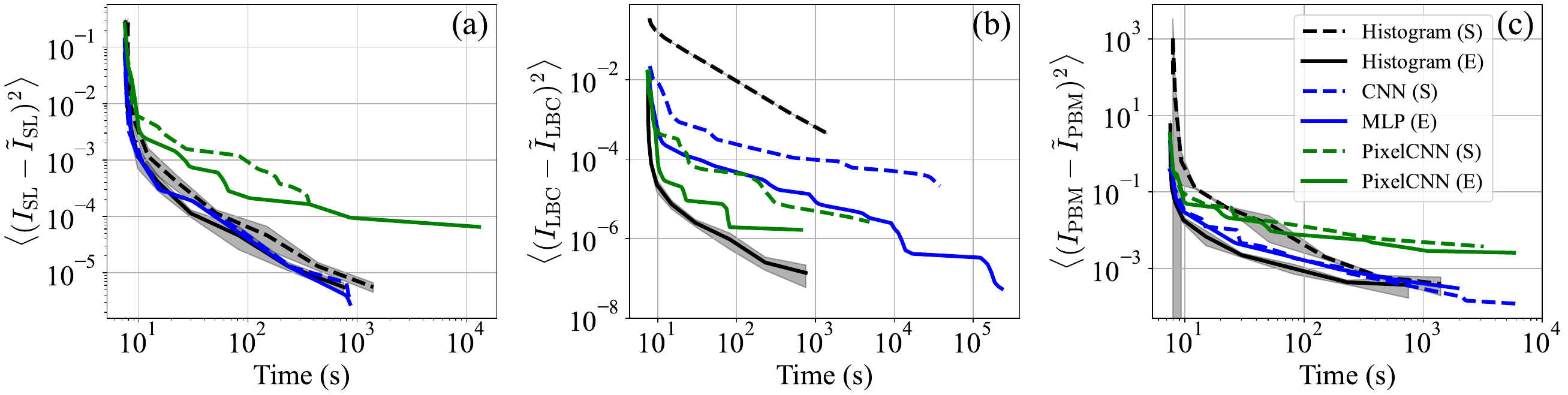}
\centering
\caption{Mean-square-error between machine-learned indicator $\tilde{I}$ and ground-truth indicator $I$ as a function of the computation time for the Ising model [Eq.~\eqref{eq:IsingH}, $L=4$] in the knowledge-driven setting. Here, the time required for data generation is included in the overall computation time. We consider the indicators of (a) supervised learning (SL), (b) learning by confusion (LBC; $l=5$), and (c) prediction-based method (PBM). Shaded bands correspond to the 1-standard deviation error estimated from 5 independent runs for the histogram-based approach. For the NN-based approaches, the envelopes of variations over hyperparameters are shown (cf. bold lines in Fig.~\ref{fig:knowledgedriven_error_comptime_complete}).}
\label{fig:knowledgedriven_error_comptime_comparison}
\end{figure}

\begin{figure}[htb!]
    \centering
    \includegraphics[width=\linewidth]{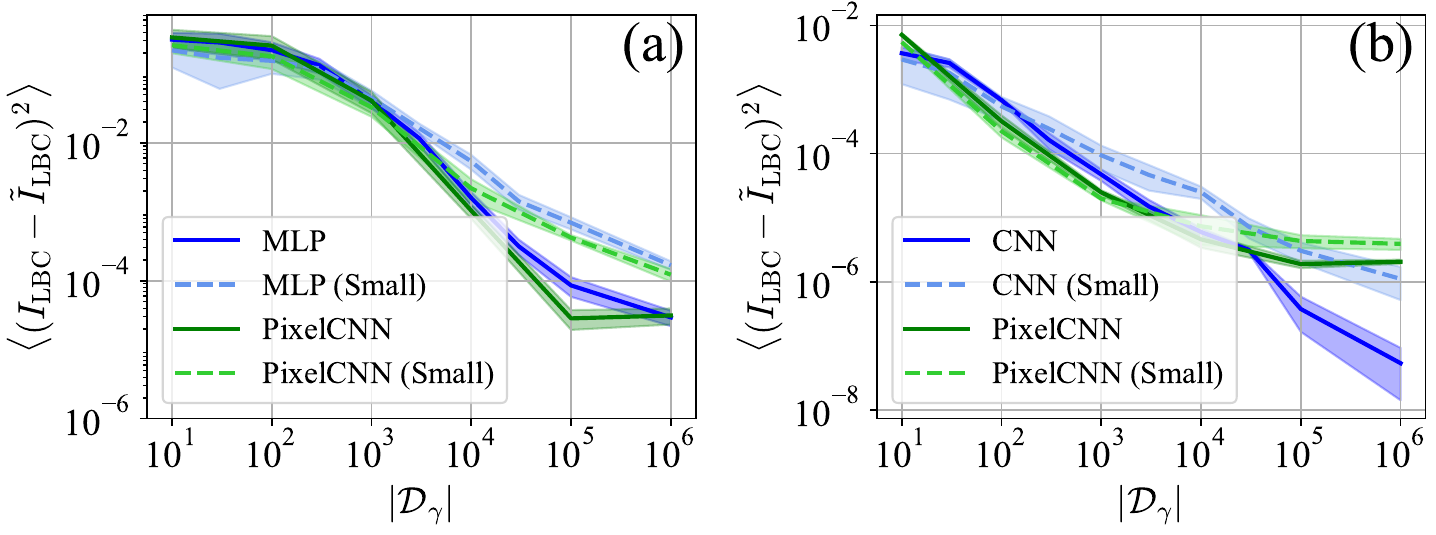}
    \caption{Mean-square-error between machine-learned indicator $\tilde{I}_{\rm LBC}$ ($l=5$) and ground-truth indicator $I_{\rm LBC}$ as a function of the dataset size for the Ising model [Eq.~\eqref{eq:IsingH}, $L=4$] in (a) the data-driven and (b) the knowledge-driven setting. For each NN-based approach we consider a smaller model variation with fewer trainable parameters. In panel (a), the small PixelCNN contained 787 trainable parameters (in contrast to the 2017  parameters of the large model) and the small MLP has 610 (in contrast to the 2242 parameters of the large model). In panel (b), the small PixelCNN has 787 trainable parameters (in contrast to the 2017 parameters of the large model) and the small MLP has 610 (in contrast to the 2242 parameters of the large model). For other hyperparameter settings, see Tabs.~\ref{tab:nn_descriptions_discriminative} and~\ref{tab:nn_descriptions_generative}. Shaded bands correspond to the 1-standard deviation error estimated from 5 independent runs for the histogram-based approach.}
    \label{fig:smaller_model_LBC}
\end{figure}

In this Appendix, we show additional results supporting our claims in the main text. Figure~\ref{fig:pixelcnn_reproduce} shows that the PixelCNNs trained both with forward and reverse KL are capable of accurately modeling the Boltzmann distribution underlying the system. In particular, the generative models are capable of reproducing key physical quantities, such as the system's average energy or magnetization. Figure~\ref{fig:knowledgedriven_error_datasetsize} shows how the error in the indicator scales as a function of the utilized dataset size within the knowledge-driven setting. Similarly, Fig.~\ref{fig:knowledgedriven_error_comptime_comparison} contains the results of all possible approaches to detect phase transitions within the knowledge-driven setting as a function of the computation time. Finally, in Fig.~\ref{fig:smaller_model_LBC} we show how the error in the LBC indicator scales as a function of the dataset size when different-sized NN as employed for both generative and discriminative models.

\clearpage  % This flushes all figures to appear before the references
%\FloatBarrier  % Use this if you're using the `placeins` package for more control

\section*{References}
\bibliography{References}
\end{document}